\documentclass[10pt, aps, prl, reprint, superscriptaddress, longbibliography]{revtex4-1} 
\usepackage[utf8]{inputenc}
\usepackage{dcolumn} 
\usepackage{bm} 
\usepackage{amsmath} 
\usepackage{amssymb}
\usepackage{xr}
\usepackage{graphicx}
\usepackage{subfigure} 
\usepackage{overpic}
\usepackage{gensymb} 
\usepackage{natbib}
\usepackage{url}
\usepackage{textcase}
\usepackage{amsfonts}
\usepackage{lineno}
\usepackage{color}
\usepackage{xcolor} 
\usepackage{amsmath} 
\usepackage{siunitx}
\usepackage{verbatim} 

\begin{document}


\title[]{(Sub-)picosecond surface correlations of femtosecond laser excited Al-coated multilayers observed by grazing-incidence x-ray scattering}

\author{L. Randolph}
\email{Lisa.Randolph@xfel.eu}
 \affiliation{Department Physik, University Siegen, 57072 Siegen, Germany}%
 \affiliation{European XFEL, 22869 Schenefeld, Germany}
\author{M. Banjafar}%
\affiliation{European XFEL, 22869 Schenefeld, Germany}%
\affiliation{Technical University Dresden, 01069 Dresden, Germany}%

\author{T. Yabuuchi}%
\affiliation{Japan Synchrotron Radiation Research Institute (JASRI)}%
\affiliation{RIKEN SPring-8 Center, Sayo, Hyogo 679-5148, Japan}%

\author{C. Baehtz}%
\affiliation{Helmholtz-Zentrum Dresden-Rossendorf, 01328 Dresden, Germany}%


\author{M. Bussmann}
\affiliation{Helmholtz-Zentrum Dresden-Rossendorf, 01328 Dresden, Germany}%
\affiliation{Center for Advanced Systems Understanding (CASUS), 02826 Görlitz, Germany}

\author{N. P. Dover}%
\affiliation{The John Adams Institute for Accelerator Science, Imperial College London, London SW7 2BW, United Kingdom}%


\author{L. Huang}
\affiliation{Helmholtz-Zentrum Dresden-Rossendorf, 01328 Dresden, Germany}%

\author{Y. Inubushi}
\affiliation{Japan Synchrotron Radiation Research Institute (JASRI)}%
\affiliation{RIKEN SPring-8 Center, Sayo, Hyogo 679-5148, Japan}%

\author{G. Jakob}%
\affiliation{Institute of Physics, Johannes Gutenberg-University, 55099 Mainz, Germany}%

\author{M. Kläui}%
\affiliation{Institute of Physics, Johannes Gutenberg-University, 55099 Mainz, Germany}%



\author{D. Ksenzov}
\affiliation{Department Physik, University Siegen, 57072 Siegen, Germany}%

\author{M. Makita}%
\affiliation{European XFEL, 22869 Schenefeld, Germany}%

\author{K. Miyanishi}
\affiliation{RIKEN SPring-8 Center, Sayo, Hyogo 679-5148, Japan}%

\author{M. Nishiushi}%
\affiliation{Kansai Photon Science Institute, National Institutes for Quantum Science and Technology, Kyoto 619-0215, Japan}%

\author{Ö. Öztürk}
\affiliation{Department Physik, University Siegen, 57072 Siegen, Germany}%

\author{M. Paulus}%
\affiliation{Fakultät Physik / DELTA, TU Dortmund, 44221 Dortmund, Germany}%

\author{A. Pelka}
\affiliation{Helmholtz-Zentrum Dresden-Rossendorf, 01328 Dresden, Germany}%

\author{T. R. Preston}%
\affiliation{European XFEL, 22869 Schenefeld, Germany}%



\author{J.-P. Schwinkendorf}%
\affiliation{European XFEL, 22869 Schenefeld, Germany}%
\affiliation{Helmholtz-Zentrum Dresden-Rossendorf, 01328 Dresden, Germany}%


\author{K. Sueda}
\affiliation{RIKEN SPring-8 Center, Sayo, Hyogo 679-5148, Japan}

\author{T. Togashi}
\affiliation{Japan Synchrotron Radiation Research Institute (JASRI)}%
\affiliation{RIKEN SPring-8 Center, Sayo, Hyogo 679-5148, Japan}%

\author{T. E. Cowan}
\affiliation{Helmholtz-Zentrum Dresden-Rossendorf, 01328 Dresden, Germany}%

\author{T. Kluge}
\affiliation{Helmholtz-Zentrum Dresden-Rossendorf, 01328 Dresden, Germany}%

\author{C. Gutt}
\affiliation{Department Physik, University Siegen, 57072 Siegen, Germany}%

\author{M. Nakatsutsumi}%
\email{Motoaki.Nakatsutsumi@xfel.eu}
\affiliation{European XFEL, 22869 Schenefeld, Germany}%

\date{\today}

\begin{abstract}
Femtosecond high-intensity laser pulses at intensities surpassing $\SI{e14}{\watt\per\centi\meter^2}$ can generate a diverse range of functional surface nanostructures. Achieving precise control over the production of these functional structures necessitates a thorough understanding of the surface morphology dynamics with nanometer-scale spatial resolution and picosecond-scale temporal resolution.
In this study, we show that individual XFEL pulses can elucidate structural changes on surfaces induced by laser-generated plasmas, employing grazing-incidence small-angle x-ray scattering (GISAXS). Using aluminum-coated multilayer samples we can differentiate between ultrafast surface morphology dynamics and subsequent subsurface density dynamics, achieving nanometer-depth sensitivity and subpicosecond temporal resolution. The observed subsurface density dynamics serve to validate advanced simulation models depicting matter under extreme conditions. Our findings promise to unveil novel avenues for laser material nanoprocessing and high-energy-density science.\\ \\
\textbf{Keywords:} grazing-incidence x-ray scattering; ultrafast surface dynamics; laser processing; XFEL
\end{abstract}
\maketitle

\section{\label{sec:Introduction}Introduction}
Intense, ultrashort laser-solid interactions at intensities ranging from $\SI{e13}{}$ to $\SI{e16}{\watt\per\centi\meter^2}$ are of paramount importance for laser nanoprocessing to achieve functional surfaces. Among these interactions, one of the most notable examples is the generation of Laser-Induced Periodic Surface Structures (LIPSS) \cite{Bonse20}, which find applications, e.g., in antibacterial coatings, optical devices, chemical sensing, and tribology ~\cite{Lutey2018, *San-Blas_LIPSS_20, *Li_LIPSS_20, *Bonse_Tribology18}.  
Understanding the dynamics of both surface and sub-surface phenomena at nanometer or even atomic scales, within the requisite temporal frame, is pivotal for comprehending the underlying physics responsible for creating desired surface structures in a controlled manner. Despite notable advancements in theoretical frameworks and models in recent years, aimed at elucidating intricate mechanisms involved in the self-organization of nanostructures under ultrashort laser irradiation within relevant temporal scales \cite{Rudenko20}, direct experimental visualization of this process remains scarce. Thus far, the majority of experimental findings regarding surface manufacturing have relied on post-mortem analyses, which lack temporal dynamics. Time-resolved experiments employing optical lasers, such as optical reflectometry, interferometry, and spectroscopy, suffer from limited spatial resolution and a lack of bulk sensitivity.
Laser processing involves a complex chain of various physical processes occurring across different temporal and spatial scales. Initially, ultrashort laser interactions with metals excite electrons within the surface skin layer, typically spanning tens of nanometers. These excited electrons then propagate into the bulk at Fermi velocity before rapid thermalization occurs within sub-ps timescales via collisions. Depending on the excitation strength and material properties, this process triggers electron-lattice/ion thermalization, coherent phonon oscillation, and lattice heating in picosecond timescales. This can lead to subsequent thermal or non-thermal melting, followed by ablation through spallation in sub-ns timescales. These phenomena collectively contribute to later crystallization or amorphization, leading to specific surface nanostructures. Therefore, understanding above-mentioned early time phenomena in (sub-)picosecond dynamics is crucial. 
To address this need, here, we demonstrate experimental capabilities of visualizing physical processes at both surface and subsurface levels with picosecond and nanometer resolutions using an X-ray Free Electron Laser (XFEL) operating in a grazing-incidence geometry.
Grazing-incidence X-ray small-angle scattering (GISAXS) is a well-established technique for probing lateral structures and correlations along the surface normal. It has been extensively utilized over the past decades at synchrotron x-ray facilities to characterize material structures across length scales ranging from sub-nanometers to microns  \cite{Roth16, Su17}, albeit limited to the millisecond timescale due to photon accumulation requirements. \\
Recently, we demonstrated the applicability of GISAXS to XFEL to track nanometric multilayer dynamics at unprecedented picosecond scales, which is six orders of magnitude faster than previously achievable \cite{Randolph22}. This breakthrough is enabled by the XFEL's ability to deliver an immense number of photons, equivalent to those typically accumulated over a second at third-generation synchrotron x-rays, within a single pulse lasting a few tens of femtoseconds. Despite inherent time smearing due to grazing-incidence, providing different parts of the x-ray arriving at the surface at different times, achieving approximately picosecond time resolution is feasible. \\
Utilizing the in-plane scattering signal, characterized by the wavevector transfer along the sample depth ($Q_z$), which correlates closely with specular reflectivity \cite{Holy93}, our experiments revealed the compressed, heated, intermixed, and ablated multilayer dynamics after the laser irradiation.
On the other hand, out-of-plane diffuse scattering along the $Q_y$ direction provides insights into ultrafast changes in the lateral distribution of surface roughnesses. 
This can be described with the help of the height-height correlation function
\begin{align}
    C(R)=<h(0)h(R)>=\sigma^2\exp{-(R/\xi_\parallel)^{2H}}
\end{align}
where $R$ is the spatial separation of two points, $\sigma$ is the rms roughness, $\xi_\parallel$ is the lateral correlation length and $H$ the hurst parameter \cite{PhysRevB.38.2297}. The correlation function along the sample depth can be written as
\begin{align}
    <h_j(0)h_k(R)>=&\frac{1}{2}\left[\frac{\sigma_k}{\sigma_j} C_j(R)+\frac{\sigma_j}{\sigma_k}C_k(R)\right]\nonumber \\
    &\times\exp{\left(-\frac{z_j-z_k}{\xi_\perp}\right)}.
\end{align} Here, $C(R)$ is the height-height correlation function, $z$ denotes the depth inside the sample and $\xi_\perp$ the cross-correlation length. The indices $j$ and $k$ denote different interfaces of the ML sample \cite{PhysRevB.51.2311}.\\

Additionally, by incorporating a low-Z capping layer on top of a high-Z multilayer (ML), GISAXS facilitates measurements of energy transport several hundred nanometers within the material. Upon ultrashort laser irradiation, strongly confined temperature gradients induce compression (shock) waves, where the progression of temperature gradients depends on the mean-free-path of excited electrons which is a function of both temperature and density. Our result show that increasing laser excitation strength transitions the dominant energy transfer mode from compression wave dominance to the electron-thermal wave dominant regime.

\section{Experimental methods and results} 
\begin{figure*}[hbt!]
    \centering
    \includegraphics[width=2\columnwidth]{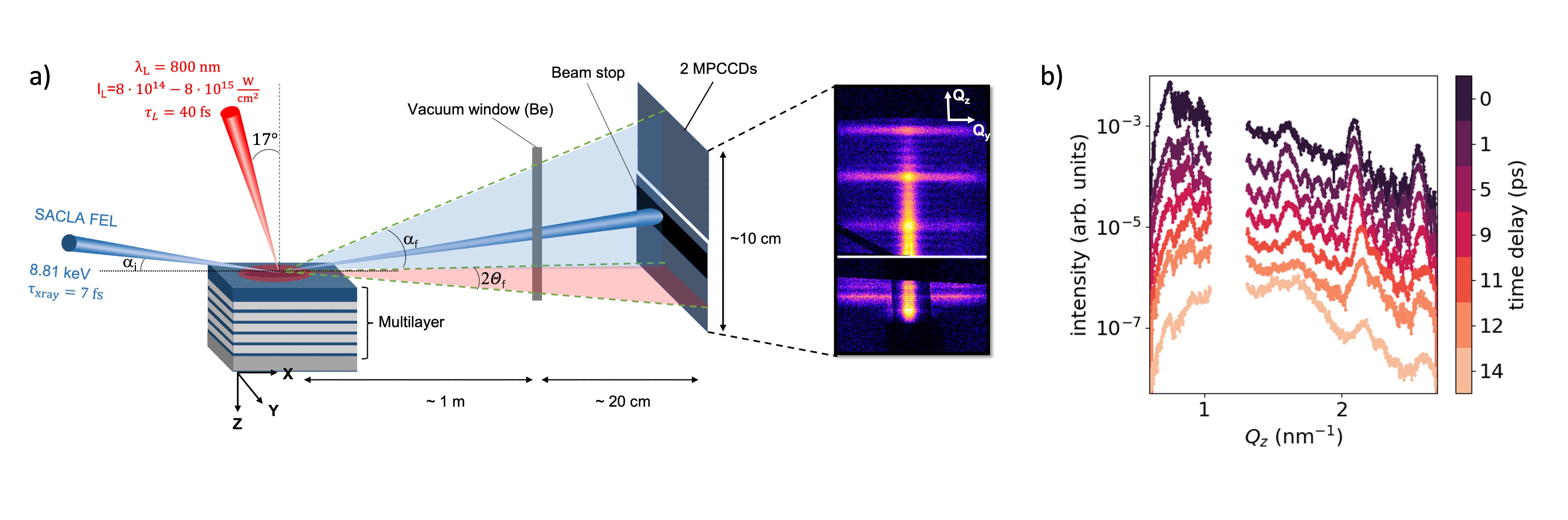}
    \caption{\label{fig:1} Schematics of the experimental setup to investigate the surface and subsurface solid-density plasma dynamics with GISAXS using single femtosecond x-ray FEL pulses. (a) A ML sample consists of 5 repetitions of tantalum (Ta) and copper nitride (Cu$_3$N) of $4.5$ and $\SI{8.5}{\nano\meter}$ thickness covered by a $\SI{200}{\nano\meter}$ aluminium capping layer. The samples were irradiated by an optical laser with a central wavelength of $\SI{800}{\nano\meter}$, intensity of $(8\pm 0.2)\times\SI{e14}{}$ and $(8\pm 0.2)\times\SI{e15}{\watt\per\centi\meter^2}$ with $\SI{40}{\femto\second}$ pulse duration. The laser is irradiated by an incident angle of $\SI{17}{\degree}$ from the surface normal in $p$ polarization. The x-ray pulses with $\SI{8.81}{\kilo\electronvolt}$ photon energy, $\SI{7}{\femto\second}$ FWHM duration are 
    irradiated on the sample at the grazing-incidence angle of $\alpha_i=\SI{0.75}{\degree}$, i.e., slightly above the critical angle of external total reflection of the layer materials. The laser beam is defocused to obtain a $\sim\SI{500}{\micro\meter}$ spot diameter to cover the x-ray footprint of $\sim\SI{300}{\micro\meter}$ at the surface. Scattered photons are recorded by 2 modules of the MPCCD area detector placed around the specular direction. The strong specular peak at $Q_z=\SI{1.16}{\nano\meter^{-1}}$ is blocked. (b) In-plane signal along $Q_z$ (at $Q_y=0$) for different time delays between $0$ and $\SI{14}{ps}$ after the laser intensity peak.}
\end{figure*}

The experiment was performed at the BL2 EH6 station at the SACLA XFEL facility in Japan \cite{Yabuuchi19}. The basic setup is similar to \cite{Randolph22}. A metallic ML sample, consisting of $5$ repetitions of tantalum (Ta, $\SI{4.5}{\nano\meter}$) and copper nitride (Cu$_3$N, $\SI{8.5}{\nano\meter}$) capped with a $\SI{200}{\nano\meter}$ thick Aluminum top layer was irradiated by an optical laser with a central wavelength of $\SI{800}{\nano\meter}$, intensities of $\SI{8e14}{}$ and $\SI{8e15}{\watt\per\centi\meter^2}$, pulse duration of $\SI{40}{\femto\second}$ and a focus size of $\sim\SI{500}{\micro\meter}$ (Fig. 1a). After a variable delay time, the surface state of the sample was probed via surface sensitive x-ray scattering employing ultrafast XFEL pulses of x-ray photons of energy $\SI{8.81}{\kilo\electronvolt}$. The x-ray beam with a pulse energy of $\sim\SI{0.1}{\milli\joule\per pulse}$ was incident under a grazing incidence angle of $\SI{0.75}{\deg}$ and the diffusely scattered intensity was recorded by a multi-port CCD (MPCCD) area detector \cite{Kameshima14} with the intense specular peak (incident angle equals exit angle, Q$_\text{specular}=\SI{1.16}{\nano\meter^{-1}}$) being blocked by a beam stop. Fig. 1 a) displays a typical single pulse GISAXS pattern. The shallow angle-of-incidence, in combination with the beam size of $\SI{4}{\micro\meter}$ leads to a large footprint ($\sim\SI{300}{\micro\meter}$) of the beam on the sample surface, limiting the temporal resolution of the detected x-ray signal to $\sim\SI{1}{\pico\second}$ integration time. \\

\begin{figure*}[hbt!]
    \centering
    \includegraphics[width=2\columnwidth]{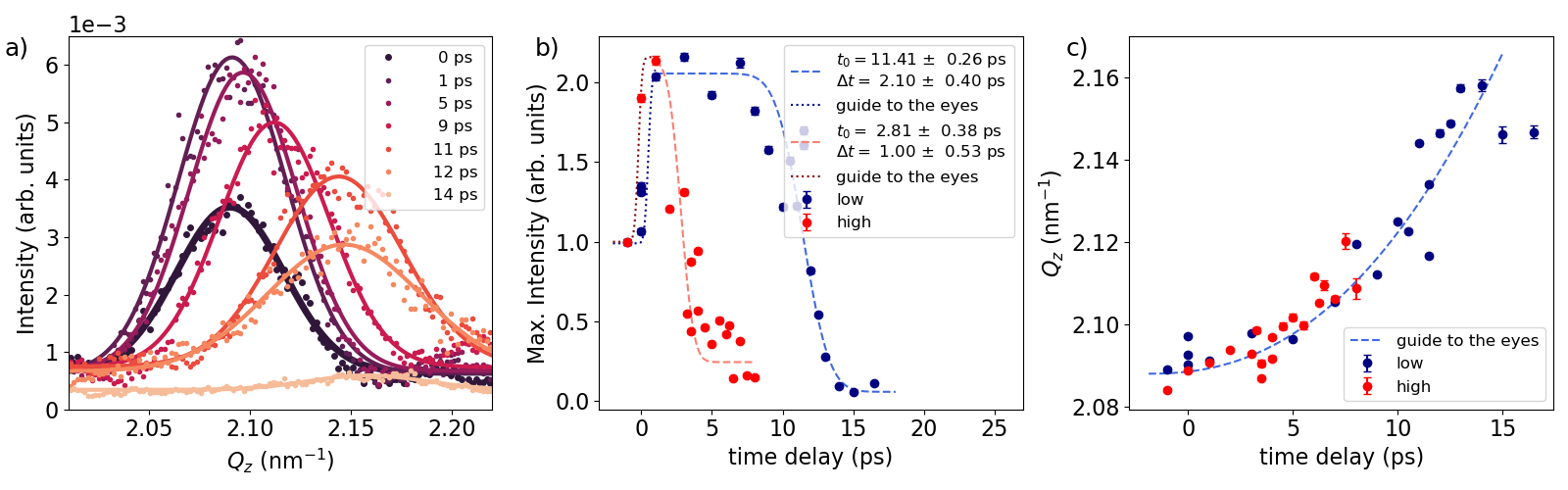}
    \caption{\label{fig:2} In-plane scattering signal along $Q_z$ around the Bragg-like peak at $\SI{2.1}{\nano\meter^{-1}}$. (a) Circular dots represent the experimental data for various time delays, while solid lines correspond to a Gaussian model fit. (b) Peak intensity derived from the Gaussian model plotted against time delay. Circular dots denote the maximum intensity values from (a), while the dashed line represents a refinement using an error function. Blue color denotes the lower laser intensity of $\SI{8e14}{\watt\per\centi\meter^2}$ while red indicates the higher laser intensity of $\SI{8e15}{\watt\per\centi\meter^2}$. (c) $Q_z$-position of the Bragg-like peak. The dashed line is a guide to the eyes.}
\end{figure*}
Fig. 1b) illustrates the in-plane scattering signal ($Q_y=0$) as a function of $Q_z$ for various delay times after laser irradiation. Because of the peculiar geometry of GISAXS, the in-plane scattering signal contains momentum transfers with components in both the normal ($Q_z$) and surface-parallel ($Q_x$) directions. As a result, the scattering signal reflects vertical density correlations along the z-direction ($Q_z$) of the ML structure, roughness correlations along the surface plane ($Q_x$) direction, and cross-correlations between different interfaces. For clarity, only the $Q_z$ values are plotted here, as $Q_x\ll Q_z$. The in-plane signal at $Q_z>Q_\text{specular}$ is closely associated with the specular reflectivity curve \cite{Holy93}, which we characterized \textit{ex-situ}. The spacing of the intense Bragg-like peaks at $Q_z=\SI{1.6}{}$, $\SI{2.1}{}$ and $\SI{2.6}{\nano\meter^{-1}}$ corresponds to the typical length scale, i.e. $\SI{13}{\nano\meter}$ thickness of each Ta/Cu$_3$N double layer. The Kiessig fringes \cite{Kiessig31}, represented as smaller peaks between the Bragg peaks, are a fingerprint of the number of double-layer repeats in the sample. Upon laser irradiation, substantial changes in the in-plane scattering signal are observed during the first $\SI{12}{\pico\second}$. The number and intensity of Kiessig fringes remain relatively stable during the initial $\sim \SI{9}{\pico\second}$ after laser irradiation, thereafter exhibiting a gradual decrease, concomitant with a simultaneous broadening and reduction of the intense Bragg-like peaks. After $\SI{12}{\pico\second}$, the Kiessig fringes are nearly gone, leaving behind only a broad residual of the Bragg-like peaks.\\

The most notable structural changes inside the ML sample are manifested by the change of intensity and position of the intense peaks at $Q_z=1.6$, $2.1$ and $\SI{2.6}{\nano\meter^{-1}}$. Fig. 2a) illustrates the temporal evolution of the peak at $Q_z=\SI{2.1}{\nano\meter^{-1}}$ following laser irradiation with an intensity of $\SI{8e14}{\watt\per\centi\meter^2}$. While only the peak at $Q_z=\SI{2.1}{\nano\meter^{-1}}$ is displayed here, qualitatively similar behaviour is observed for the peaks at $Q_z=\SI{1.6}{}$ and $\SI{2.6}{\nano\meter^{-1}}$. The solid lines represent refinements using a Gaussian function. The peak amplitude as a function of time delay, extracted from the Gaussian refinement, is depicted in Fig. 2b), with blue dots, which shows an ultrafast increase within the first picosecond. Subsequently, the intensity remains constant, followed by a gradual decrease at around $\sim\SI{10}{\pico\second}$ associated with a loss of structural correlation between the layers. This decay is modelled using an error function 
\begin{equation}
    \text{Intensity}=A\cdot\text{erf}\left(-\frac{(x-t_0)}{\Delta t}\right)-B
\end{equation}
resulting in values for $t_0=11.4\pm\SI{0.3}{\pico\second}$ and a width of $\Delta t =2.1\pm \SI{0.4}{\pico\second}$. Here, $A$ represents a stretching factor, $t_0$ denotes the time at which the error function decreased to half of its initial value, $\Delta t$ is the duration of the decrease, and $B$ stands for a constant offset.  In comparison, the same analysis for the higher laser intensity case of $\SI{8e15}{\watt\per\centi\meter^2}$, is depicted with red circular dots. Similar to the lower intensity case, the ultrafast increase in scattering intensity within the first $\sim\SI{1}{\pico\second}$ is evident. However, for the time delays $>\sim\SI{1}{\pico\second}$ the amplitude quickly begins to decline again. The refined error function yields a time constant of $t_0=2.8\pm \SI{0.3}{\pico\second}$ and a width of $\Delta t=1.0\pm \SI{0.5}{\pico\second}$. 
On top of the change in the Bragg intensity, we observe a steady increase in the Q-position with time delay for both laser intensities, as summarized in Fig. \ref{fig:2} c). This shift towards larger Q-values implies compression of the double-layer structure as the typical length scale decreases. \\
\begin{figure}[hbt!]
    \centering
    \includegraphics[width=1\columnwidth]{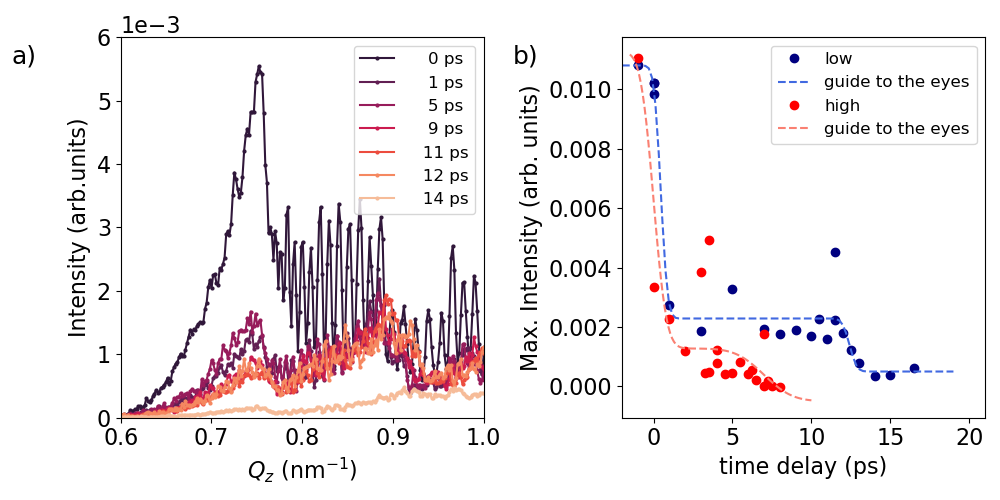}
    \caption{\label{fig:3} In-plane signal along $Q_z$ around the dynamical diffraction area between $\SI{0.6}{}$ and $\SI{1.0}{\nano\meter^{-1}}$. (a) In-plane signal for different time delays at lower laser intensity case ($\SI{8e14}{\watt\per\centi\meter^2}$). (b) The circular dots display the integrated signal between $Q_z=0.65-\SI{0.79}{\nano\meter^{-1}}$ as a function of time delay. The dashed line represents a guide to the eyes. Blue color stands for the lower laser intensity of $\SI{8e14}{\watt\per\centi\meter^2}$ and red denote the higher laser intensity $\SI{8e15}{\watt\per\centi\meter^2}$.}
\end{figure}

The dynamic diffraction effects at $Q_z<Q_\text{specular}$ yield insight into the dynamics of the top Al layer. In particular, the scattering signals at exit angles equal to the critical angle for the total external reflection -- $Q_z$ = $\SI{0.75}{\nano\meter^{-1}}$ for Al -- are particularly surface-sensitive, originating from an evanescent x-ray wave travelling parallel to the surface. 
This so-called Yoneda peak \cite{Yoneda63}, primarily originated from interference within the topmost surface layer, serves as a sensitive marker of its surface structure. A close-up of this Yoneda peak is summarized in Fig. \ref{fig:3} a). The peak at $Q_z=\SI{0.75}{\nano\meter^{-1}}$ exhibits a progressive decrease after laser irradiation, which persists for at least at $t = \SI{12}{\pico\second}$ before almost disappearing at $t = \SI{14}{\pico\second}$, indicative of the ongoing presence of the solid density Al cover layer. Assuming the surface ablation speed approximates the speed of sound, $C_s = \sqrt{{(\gamma_e Z_{mean} k_B T_e + \gamma_i k_B T_i)}/{M_i}}$, the ablation of the $\SI{200}{\nano\meter}$ layer would require $\sim\SI{15}{\pico\second}$. Here, $T_e \sim 10 \mathrm{eV}$ is the electron temperature, $T_i \sim 1 \mathrm{eV}$ is the ion temperature, $Z_{mean} \sim 5$ is the mean ionization, $\gamma_e = 1$ and $\gamma_i = 3$ are adiabatic index of electrons and ions, respectively and $M_i=26$ a.u. is the ion mass for Al. \\ The integrated intensity of the Yoneda peak ($Q_z=0.65-\SI{0.79}{\nano\meter^{-1}}$) is depicted in Fig. \ref{fig:3} b) for both low (blue) and high (red) intensity cases. In both cases, a rapid initial decay in intensity is observed within the first picosecond, followed by a period of quasi-constant signal intensity extended up to $\sim\SI{12}{\pico\second}$ and $\sim\SI{5}{\pico\second}$ before subsequent decrease, respectively. This temporal behaviour aligns qualitatively with that observed at the Bragg-like peak at higher $Q_z$ shown in Fig. \ref{fig:2} b). 
In Fig. \ref{fig:3} a), two additional Yoneda peaks are also visible at higher $Q_z$, corresponding to the Yoneda peaks for $\mathrm{Cu_3N}$ ($Q_z = \SI{0.86}{\nano\meter^{-1}}$) and Ta ($\SI{0.95}{\nano\meter^{-1}}$), respectively. Given that these materials are embedded within a $\SI{200}{\nano\meter}$ thick Al layer, the time-dependency of their intensity is not as pronounced as that of the Al Yoneda peak, except during the period between 12 and $\SI{14}{\pico\second}$ when all peaks suddenly disappear. As we will discuss below, aided by plasma simulations, this phenomenon can be attributed to the arrival of a compression wave to the ML, which initiates significant modulation of the ML structure. 
An additional noteworthy feature in the figure is the presence of high-frequency fringes superimposed on the Ta and $\mathrm{Cu_3N}$ Yoneda regions. These fringes arise from the interference between waves scattered from the Al surface and those from the Al-ML interface. Following laser excitation and the subsequent expansion of the top Al surface, these fringes rapidly vanish as a result of blurring of the distinct interface between vacuum and Al.  
 \\

\begin{figure}[hbt!]
    \centering
    \includegraphics[width=1\columnwidth]{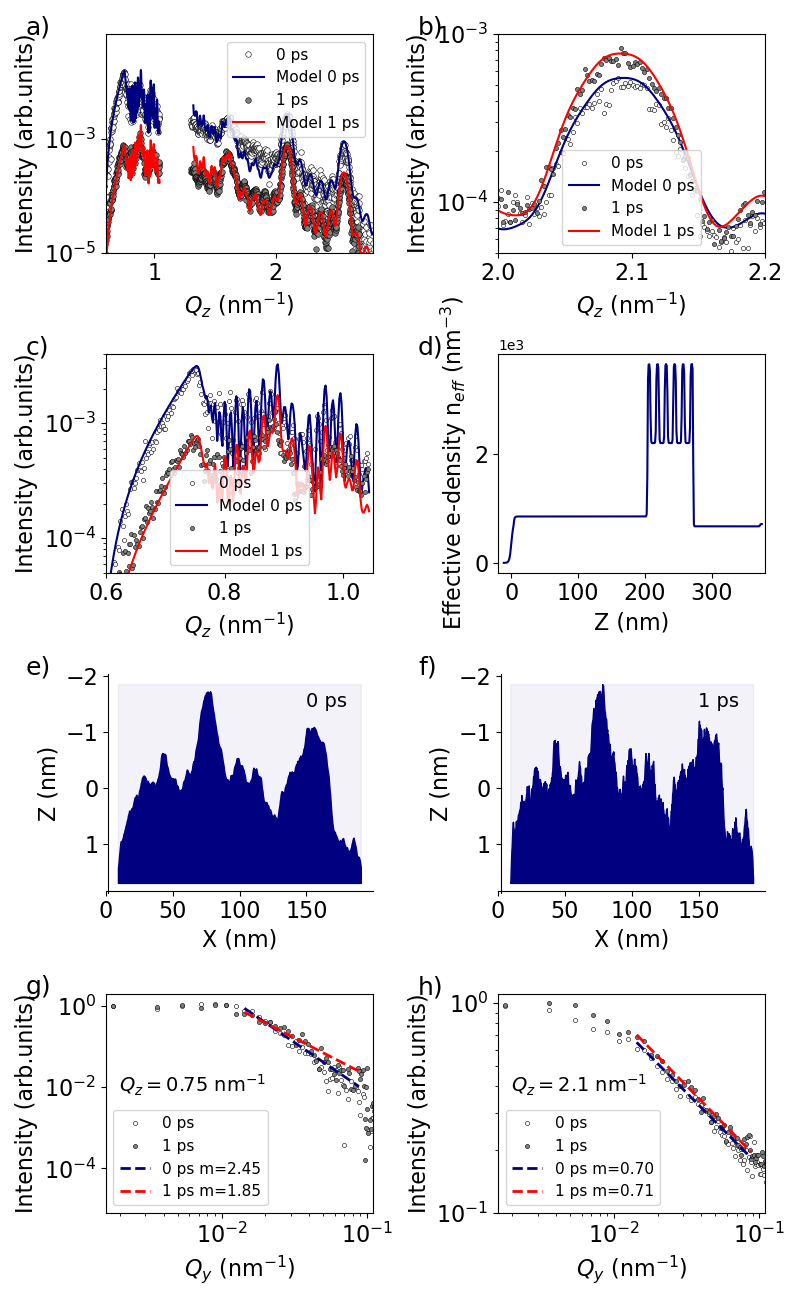}
    \caption{GISAXS signals and corresponding retrieved real-space electron density profile. (a) Lineout at $\SI{0}{\pico\second}$ (white circular dots) and $\SI{1}{\pico\second}$ (grey circular dots) after the laser intensity peak. Solid lines represent models using the program BornAgain. (b) Zoom into the area of the Bragg-like peak at $Q_z=\SI{2.1}{\nano\meter^{-1}}$ and (c) zoom into the area of dynamical diffraction around $Q_z=\SI{0.75}{\nano\meter^{-1}}$. (d) Retrieved real-space effective electron density profile as a function of the depth (Z). The laser irradiates the sample from the left. (e) Generated surface model for $H=0.6$, $\sigma=\SI{2.3}{\nano\meter}$ and (f) $H=0.15$, $\sigma=\SI{2.6}{\nano\meter}$.  (g) Lineout along $Q_y$ at the Al Yoneda peak ($Q_z=\SI{0.75}{\nano\meter^{-1}}$). Dashed lines indicate a fit of the decay. (h) Lineout along $Q_y$ at $Q_z=\SI{2.1}{\nano\meter^{-1}}$.}
\end{figure}

Both Bragg-like peaks and the Yoneda peak indicate the presence of three distinct time regimes. Immediately after laser irradiation $\sim \SI{1}{\pico\second}$, we observe an ultrafast increase in the intensity of the Bragg-like peaks without any change in their $Q_z$-positions, accompanied by the decay of the Yoneda peak. 
As we will discuss further below, this phenomenon is attributed to a modification in the surface structural properties. 
In the second time frame, spanning from $1$ to $\sim\SI{10}{\pico\second}$ or to $\sim\SI{3}{\pico\second}$ for low and high laser intensities, respectively, the intensities of the Bragg-like peaks and Yoneda peak remain constant, while the $Q_z$ position of the Bragg-like peaks steadily increases. This suggests that the embedded ML undergoes compression, although ML structure remains constant. 
In the third time frame, the intensities of the Bragg and Yoneda peaks are further reduced, accompanied by a continued shift of the $Q_z$-position to even higher values. This indicates further compression of the entire multilayer, as well as significant modulation of the ML structure, resulting in the loss of correlation and subsequent decrease in the x-ray scattering intensity. 

\section{Discussion} 
To facilitate a more quantitative discussion of the ultrafast surface dynamics, we employed the BornAgain \cite{Pospelov:ge5067} GISAXS analysis program to model our experimental observations. The white circular dots in Fig. 4a) represent the experimental data at $\SI{0}{\pico\second}$ while the grey dots depict the signal at a delay of $\SI{1}{\pico\second}$. The solid lines represent the BornAgain model corresponding to the effective electron density profiles shown in Fig. 4d). Figures 4b) and c) provide enlarged views of the intense peak at $\SI{2.1}{\nano\meter^{-1}}$ and the surface-sensitive Yoneda region, respectively. According to the model, the observed variation between the blue ($\SI{0}{\pico\second}$) and red ($\SI{1}{\pico\second}$) lines in Figs 4a)-c) can be attributed to the decrease in the Hurst parameter of the Al surface layer from $0.6$ to $0.15$, indicating an increase in the spatial frequency of roughness, as illustrated schematically in Figs. 4 e) and f).  
This observation is supported by analyzing the out-of-plane signal along $Q_y$ at different $Q_z$ positions. Fig. 4g) illustrates a lineout along $Q_y$ at the Al Yoneda peak. The decay starting at $Q_y = \SI{0.01}{\nano\meter^{-1}}$ is modeled via
\begin{equation}
    \text{Intensity}\sim Q_y^{-m}
\end{equation}
where $m$ defines the slope of the decay which is proportional to the Hurst parameter $H$ \cite{PhysRevLett.73.2228}. Refining the lineouts for $0$ and $1$ ps indeed reveals a reduction of the Hurst parameter as $m$ has decreased. Conversely, the lineout along $Q_y$ at larger $Q_z = \SI{2.1}{\nano\meter^{-1}}$ (Fig. 4h)) shows no change in the Hurst parameter, indicating that the ultrafast change is localized to the surface rather than at the ML interfaces.
Furthermore, it appears that in order to align the model with the experimental observation from $0$ to $\SI{1}{\pico\second}$, the vertical correlation length needs to be increased from $100$ to $\SI{150}{\nano\meter}$, and the surface roughness RMS $\sigma$ should be slightly increased from $2.3$ to $\SI{2.6}{\nano\meter}$.
We hypothesize that this ultrafast change in surface roughness properties to be attributed to the presence of a thin aluminium oxide layer on the surface, which typically forms within minutes when exposed to air \cite{doi:10.1021/acsami.7b17224}. This oxide layer likely possesses a distinct surface morphology compared to the intrinsic surface morphology parameters of bare aluminium, leading to a reduction in spatial frequency of roughness as well as a decrease in vertical correlation length. Given that this layer is typically extremely thin, on the order of a nanometer, it evaporates instantaneously ($<\SI{1}{\pico\second}$) after laser excitation revealing the intrinsic surface properties of the initial Al layer.
\\

\begin{figure*}[hbt!]
         \centering
         \begin{subfigure}{}
           \includegraphics[width=0.8\linewidth]{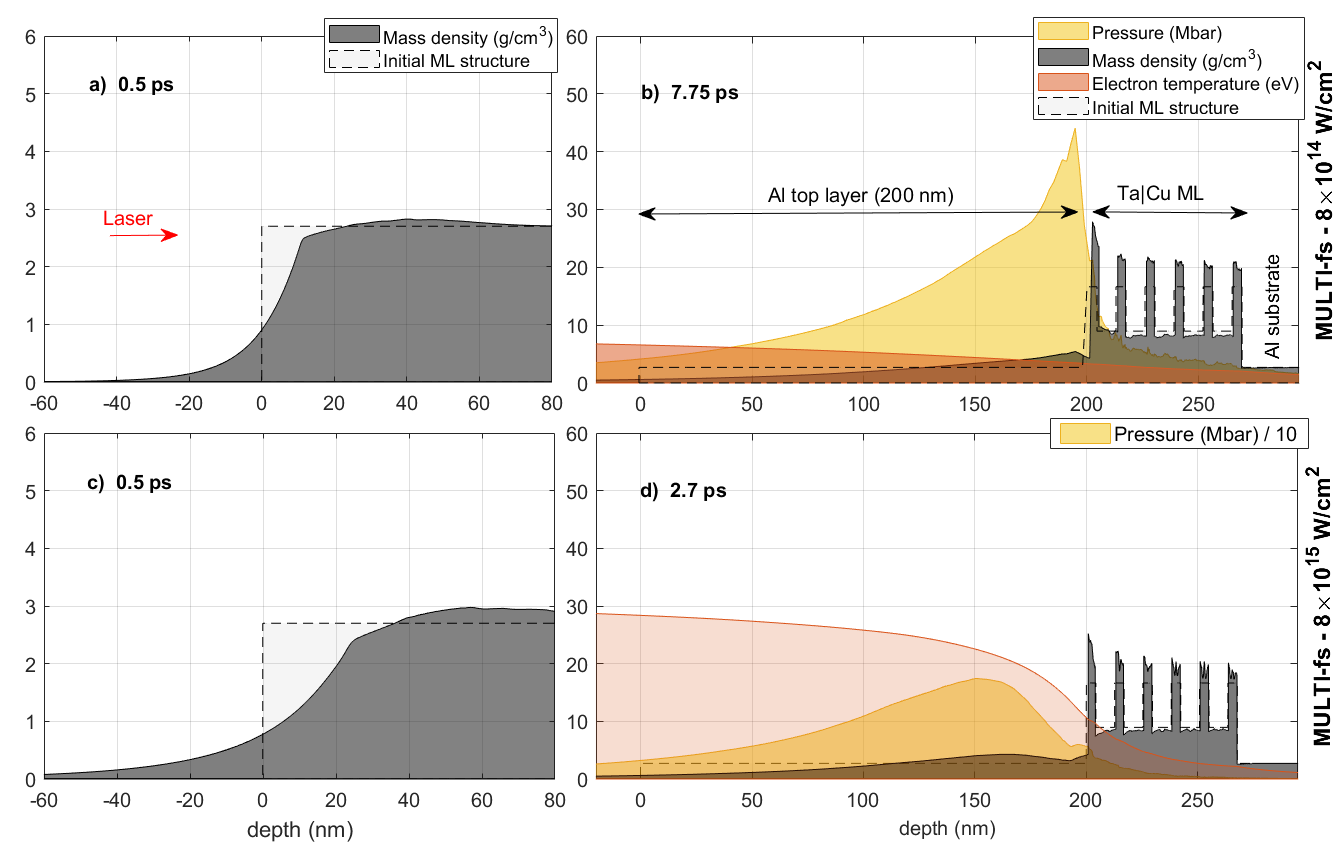}
        \end{subfigure}\\
        \vspace*{-5mm}
        \begin{subfigure}{}
           \includegraphics[width=0.8\linewidth]{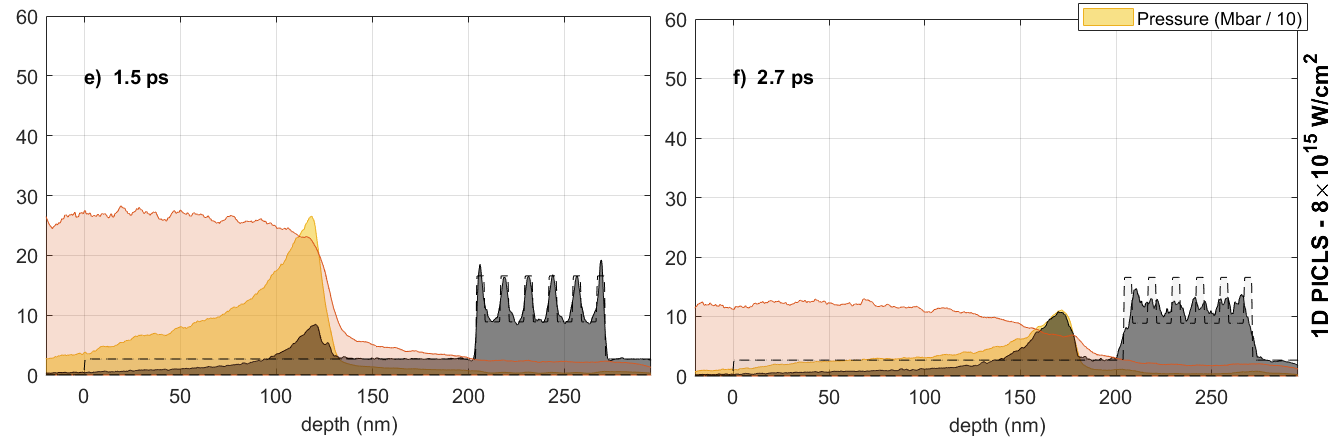}
        \end{subfigure}
         \vspace*{0mm}
         \caption{MULTI-fs hydrodynamic simulations irradiated by a p-polarized laser pulse with a duration of 50 fs, a wavelength of 800~nm, and a laser intensity of $\SI{8e14}{\watt\per\centi\meter^2}$ at time delays of (a) 0.5 and (b) 7.75 ps, and with a laser intensity of $\SI{8e15}{\watt\per\centi\meter^2}$ at time delays of (c) 0.5 and (d) 2.7 ps. The ML structure matched that of the experimental setup. (e and f) 1D Particle-in-cell PICLS simulation results at time delays of (e) 1.5 and (f) 2.7 ps. In all figures, the yellow, grey, and red regions represent the laser-induced pressure wave (in Mbar or Mbar$/10$), mass density (in $\mathrm{g/cm^3}$), and electron temperature (in eV), respectively. The dashed line denotes the initial target density. The red arrow in (a) indicates the direction of the incoming laser.}
         \label{fig:SACLA_Setup}
     \end{figure*}


To substantiate our experimental findings further, we also conducted two advanced simulations: a 1D~MULTI-fs hydrodynamic simulation and 1D~PICLS Particle-in-Cell (PIC) simulation.
The 1D~MULTI-fs hydrodynamic simulation code is tailored for modeling short pulse high-intensity laser-solid interactions which directly solves the Maxwell equations. The code incorporates a temperature-dependent collision frequency, thermal conductivity ranging from metallic solids to high-temperature ideal plasmas, and a distinct equation-of-state (EOS) for electrons and ions (two-temperature model). The details of these simulations' setup is summarized in \textit{Methods} section.
The 1D~PICLS simulation incorporates the collisional 1d3v (one-dimensional in space and three-dimensional in velocity)~\cite{SENTOKU2008}. An interpolated collision frequency akin to the MULTI-fs method was integrated into the code to address Angstrom-scale collisions occurring at electron temperatures ($T_e$) around the Fermi temperature ($T_F$). The setup details for these simulations are summarized in the \textit{Methods} section.

Figs. 5 a) and c) display the density profile simulated using MULTI-fs at a delay of 0.5 ps. The localized heating induced by the laser within the skin depth instantaneously elevates the surface electron temperature to a few tens of eV and generates pressures exceeding $>\SI{10}{\mega\bar}$, prompting immediate surface expansion. This observation is aligned with the experiment demonstrating ultrafast changes in surface properties. The resultant expansion causes an evaporation of the top few nanometers. This value contributes to the roughness RMS parameter for the GISAXS pattern. 

During the second time frame, spanning from $1$ to $\sim \SI{9}{\pico\second}$ in the case of lower laser intensity ($I_L=\SI{8e14}{\watt\per\centi\meter^2}$), the intensity of the Bragg-like peak remains constant (Fig.2 b)). As confirmed by MULTI-fs simulations, it appears that between $1$ and $\SI{8}{\pico\second}$ the compression wave is only in the Al layer until it arrives at the Al-ML boundary (Fig. 5 b), highlighted in yellow). Primarily, only the Al layer undergoes modulation during this interval, while the ML remains static with slight compression. 
Subsequently, after $\sim\SI{8}{\pico\second}$, the simulation reveals a strong pressure wave compressing the ML. This aligns with the experimental findings (Fig. 2 c)) that demonstrated a further increase in the $Q_z$-position of the Bragg peaks. 

Figures. 5 c)-d) are the same simulation but performed at the higher laser intensity $I_L=\SI{8e15}{\watt\per\centi\meter^2}$. Here, the pressure peak arrives at the multilayer surface already after $\sim$ 3~ps, which encounters strong modulation subsequently. 
On the contrary, the experiment (Fig. 2 b)) shows the significant decrease in Bragg intensity already started at $1-\SI{2}{\pico\second}$. It implies that the significant modulation of the ML occurs before the arrival of the strong compression wave to the ML surface, which is not seen in the MULTI-fs simulation. 

Lagrangian hydrodynamic simulations like MULTI-fs are incapable of accurately simulating the atomic mixing of adjacent layers \cite{Randolph22}. Therefore, we conducted kinetic 1D-PICLS simulation for the intensity case of $I_L=\SI{8e15}{\watt\per\centi\meter^2}$, summarized in Fig. 5 e-f). The simulation shows that the embedded ML undergoes significant layer intermixing from a thermal wave, even before the arrival of the strong compression wave. This phenomenon begins already at 1.5 ps delay as shown in Fig. 5 e). Even though the pressure peak is still in the middle of the Al layer, the ML has already started to undergo modulation. This leads to reductions in the Bragg peak in the x-ray scattering. By 2.7 ps delay, when both MULTI-fs and PICLS indicate the arrival of the pressure peak (Fig. 5 f)), PICLS demonstrates that the entire ML is intermixed, and the original ML structure is entirely disrupted. Such a structure cannot produce any pronounced Bragg-like peaks due to the absence of a statistically relevant periodic structure. 

In summary, for the low-intensity case $I_L=\SI{8e14}{\watt\per\centi\meter^2}$, modifications to the ML structure are primarily driven by the compression wave, as the electron temperature is low enough that the thermal wave cannot significantly alter the ML before the arrival of the pressure wave. Conversely, for the high-intensity case $I_L=\SI{8e15}{\watt\per\centi\meter^2}$, significant heating occurs deep within the ML much before the arrival of the pressure wave.



\section{Conclusions} 
In conclusion, our investigation of surface and subsurface dynamics in high-intensity laser-excited ML systems using ultrafast GISAXS has provided important insights into the complex physical processes occurring in laser-matter interactions on a picosecond time scale.\\
We can distinguish between ultrafast surface dynamics and subsequent subsurface dynamics. The ultrafast modification in surface properties occurs in sub-picosecond time scale  induced by strong localized surface heating. The slower subsurface dynamics manifest as changes in intensity and position of the intense Bragg-like peak at $Q_z=\SI{2.1}{\nano\meter^{-1}}$. Comparison of the temporal evolution of the density dynamics with 1D hydrodynamic and 1D particle-in-cell simulations reveals qualitative agreement across different time regimes observed in the experiment. Notably, the observed time scales align well with experimental observations, indicating the arrival of the pressure wave at the ML surface and subsequent intermixing of individual layers. \\
The presented GISAXS experiment was limited in time resolution to $\sim\SI{1.2}{\pico\second}$ due to the large footprint in grazing-incidence. However, this can be improved by using a smaller x-ray focal spot, e.g. $\SI{100}{\nano\meter}$, which is readily available at XFEL facilities. This would imply only $\sim\SI{30}{\femto\second}$ smearing at $\SI{0.75}{\degree}$ grazing-incidence, allowing experimental studies on surface and subsurface dynamics with nanometer spatial and femtosecond temporal resolution to observe the ultrafast changes on the surface or using laser intensities above $\SI{e16}{\watt\per\centi\meter^2}$  to study e.g. the effect of density oscillations \cite{Paschke2022}.\\

\section {Supplementary Materials}
\subsection{Multi-fs}
The 1D~MULTI-fs hydrodynamic simulation code is tailored for modeling short ($\leq$ ps) pulse high-intensity laser-solid interactions at intensities below $< 10^{17} \,\mathrm{W/cm^2}$~\cite{Eidmann00, Ramis2012}. MULTI-fs directly solves the Maxwell equations, capturing ultrashort laser plasma dynamics with sharp plasma density gradients. The code incorporates a temperature-dependent collision frequency, thermal conductivity ranging from metallic solids to high-temperature ideal plasmas, and a distinct equation-of-state (EOS) for electrons and ions (two-temperature model). 
In the MULTI-fs simulation setup, the simulation box comprises 1478 cells distributed as follows: 400 cells for the 200~nm-thick Al top layer, 678 cells for the ML, and 400 cells for a 200~nm-thick Al substrate. To mitigate numerical artifacts, finer cell resolutions were employed near layer interfaces. 
We employed the EOS and ionization tables for aluminium (Al) as provided by the MULTI-fs package. For tantalum (Ta) and copper (Cu), the EOS and ionization data were generated using the FEOS code~\cite{Faik18}.
The radiation transfer module was disabled in the simulation, and the free streaming limiting factor was set to $f=0.6$. 
To maintain consistency with the experimental setup, the laser angle of incidence was fixed at $ 17 \degree$ from the target normal. 
In our simulations, Al was chosen as the substrate material instead of silicon (Si) used in the experiment. This choice was motivated by aluminium's well-established EOS. Given that the energy transfer occurs from the ML to the substrate, we anticipate that the choice between these two materials will have minimal impact on the dynamics of the ML.

\subsection{1D~PICLS}
The 1D~PICLS simulation incorporates the collisional 1d3v (one-dimensional in space and three-dimensional in velocity)~\cite{SENTOKU2008}. An interpolated collision frequency akin to the MULTI-fs method was integrated into the code to address Angstrom-scale collisions occurring at electron temperatures ($T_e$) around the Fermi temperature ($T_F$). To accurately simulate the microscopic particle collisions at the atomic scale, the spatial resolution was configured with a cell size of $\Delta x = 2$~\AA~corresponding to a time--step of $\Delta t = 6.67 \times 10^{-19}$~s. Each computational cell accommodated 30 virtual ion particles with initial charge states of 2, 1, and 3 for Ta, Cu, and Al, respectively. Additionally, a fourth-order particle shape was employed, along with distinct particle weightings for different ion species. 
The ion number densities are set to realistic values for Ta ($n_{\mathrm{Ta}}=31.8n_c$), Cu ($n_{\mathrm{Cu}}=48.7n_c$), and Al ($n_{\mathrm{Al}}=34.6n_c$), where $n_c = m_e \omega_L^2/(4 \pi e^2) = 1.742 \times 10^{21} \mathrm{cm^{-3}}$ represents the critical plasma density at the laser wavelength of $\lambda_L = 800$~nm. Here, $m_e$ and $e$ denote the electron mass and charge, respectively, and $\omega_L$ is the laser angular frequency. Ionization dynamics are modeled employing field and direct-impact ionization models. The laser incident angle is set as normal to the surface, as oblique incidence is not supported in 1D PICLS. Our test simulations using the MULTI-fs have revealed a negligible difference, approximately 1\%, in laser absorption between normal incidence and a 17-degree angle of incidence.

\section*{Acknowledgements}
The XFEL experiments were performed at the BL2 of SACLA with the approval of the Japan Synchrotron Radiation
Research Institute (JASRI) (Proposal No. 2019B8076). We thank E. Brambrink, S. Göde, S.V. Rahul, C. Rödel, A. Kon, J. Koga and Y. Sentoku for various discussions and advices.

\subsection*{Funding}
C.G. and M.Na. acknowledge funding by DFG GU 535/6-1. G.J.  and M.K.  acknowledge the support by the Deutsche Forschungsgemeinschaft (DFG, German
Research Foundation) Project No. 268565370 (SFB TRR173 Projects A01 and B02) by TopDyn and the ForLab MagSens.

\subsection*{Author contributions}

C.G. and M.Na. conceived the study. L.R., M.Ba., T.R.P., M.M., N.P.D., S.G., T.M., M.Ni., A.P., J.-P.S., C.B., C.G., and M.Na. performed the experiment at SACLA with support from T.Y., Y.I., K.S., and T.T. L.R. analyzed the GISAXS data under the supervision of C.G. M.Ba. performed MULTI-fs and PICLS simulations to interpret the data with extensive support of L.H. and T.K, under the supervision of A.P.M., M. Bu., T.E.C., and
M.Na. N.P.D. and J.K.K. gave support for generating EOS tables. Multilayer samples were prepared by G.J. and M.K. which were further characterized by L.R. and M.P. before the experiment. J.K., M.M., and M.Na. organized the sample cutting. L.R, M.Ba, C.G. and M.Na. wrote the initial draft which was extensively revised by contribution from D.K. and Ö.Ö. All authors read commented on the manuscript.

\subsection*{Data Availability Statement}
The data are available upon reasonable request. 

\subsection*{Conflicts of Interest}
The authors declare no conflicts of interest.

\bibliography{aipsamp.bib}

\begin{thebibliography}{24}%
\makeatletter
\providecommand \@ifxundefined [1]{%
 \@ifx{#1\undefined}
}%
\providecommand \@ifnum [1]{%
 \ifnum #1\expandafter \@firstoftwo
 \else \expandafter \@secondoftwo
 \fi
}%
\providecommand \@ifx [1]{%
 \ifx #1\expandafter \@firstoftwo
 \else \expandafter \@secondoftwo
 \fi
}%
\providecommand \natexlab [1]{#1}%
\providecommand \enquote  [1]{``#1''}%
\providecommand \bibnamefont  [1]{#1}%
\providecommand \bibfnamefont [1]{#1}%
\providecommand \citenamefont [1]{#1}%
\providecommand \href@noop [0]{\@secondoftwo}%
\providecommand \href [0]{\begingroup \@sanitize@url \@href}%
\providecommand \@href[1]{\@@startlink{#1}\@@href}%
\providecommand \@@href[1]{\endgroup#1\@@endlink}%
\providecommand \@sanitize@url [0]{\catcode `\\12\catcode `\$12\catcode
  `\&12\catcode `\#12\catcode `\^12\catcode `\_12\catcode `\%12\relax}%
\providecommand \@@startlink[1]{}%
\providecommand \@@endlink[0]{}%
\providecommand \url  [0]{\begingroup\@sanitize@url \@url }%
\providecommand \@url [1]{\endgroup\@href {#1}{\urlprefix }}%
\providecommand \urlprefix  [0]{URL }%
\providecommand \Eprint [0]{\href }%
\providecommand \doibase [0]{http://dx.doi.org/}%
\providecommand \selectlanguage [0]{\@gobble}%
\providecommand \bibinfo  [0]{\@secondoftwo}%
\providecommand \bibfield  [0]{\@secondoftwo}%
\providecommand \translation [1]{[#1]}%
\providecommand \BibitemOpen [0]{}%
\providecommand \bibitemStop [0]{}%
\providecommand \bibitemNoStop [0]{.\EOS\space}%
\providecommand \EOS [0]{\spacefactor3000\relax}%
\providecommand \BibitemShut  [1]{\csname bibitem#1\endcsname}%
\let\auto@bib@innerbib\@empty
\bibitem [{\citenamefont {Bonse}\ and\ \citenamefont
  {Gr\"{a}f}(2020)}]{Bonse20}%
  \BibitemOpen
  \bibfield  {author} {\bibinfo {author} {\bibfnamefont {J.}~\bibnamefont
  {Bonse}}\ and\ \bibinfo {author} {\bibfnamefont {S.}~\bibnamefont
  {Gr\"{a}f}},\ }\bibfield  {title} {\enquote {\bibinfo {title} {{Maxwell Meets
  Marangoni -- A Review of Theories on Laser‐Induced Periodic Surface
  Structures}},}\ }\href {\doibase 10.1002/lpor.202000215} {\bibfield
  {journal} {\bibinfo  {journal} {Laser and Photonics Reviews}\ }\textbf
  {\bibinfo {volume} {14}},\ \bibinfo {pages} {2000215} (\bibinfo {year}
  {2020})}\BibitemShut {NoStop}%
\bibitem [{\citenamefont {Lutey}\ \emph {et~al.}(2018)\citenamefont {Lutey},
  \citenamefont {Gemini}, \citenamefont {Romoli}, \citenamefont {Lazzini},
  \citenamefont {Fuso}, \citenamefont {Faucon},\ and\ \citenamefont
  {Kling}}]{Lutey2018}%
  \BibitemOpen
  \bibfield  {author} {\bibinfo {author} {\bibfnamefont {Adrian~H.A.}\
  \bibnamefont {Lutey}}, \bibinfo {author} {\bibfnamefont {Laura}\ \bibnamefont
  {Gemini}}, \bibinfo {author} {\bibfnamefont {Luca}\ \bibnamefont {Romoli}},
  \bibinfo {author} {\bibfnamefont {Gianmarco}\ \bibnamefont {Lazzini}},
  \bibinfo {author} {\bibfnamefont {Francesco}\ \bibnamefont {Fuso}}, \bibinfo
  {author} {\bibfnamefont {Marc}\ \bibnamefont {Faucon}}, \ and\ \bibinfo
  {author} {\bibfnamefont {Rainer}\ \bibnamefont {Kling}},\ }\bibfield  {title}
  {\enquote {\bibinfo {title} {{Towards laser-textured antibacterial
  surfaces}},}\ }\href {\doibase 10.1038/s41598-018-28454-2} {\bibfield
  {journal} {\bibinfo  {journal} {Scientific Reports}\ }\textbf {\bibinfo
  {volume} {8}},\ \bibinfo {pages} {1--10} (\bibinfo {year}
  {2018})}\BibitemShut {NoStop}%
\bibitem [{\citenamefont {San-Blas}\ \emph {et~al.}(2020)\citenamefont
  {San-Blas}, \citenamefont {Martinez-Calderon}, \citenamefont {Buencuerpo},
  \citenamefont {Sanchez-Brea}, \citenamefont {del Hoyo}, \citenamefont
  {G{\'{o}}mez-Aranzadi}, \citenamefont {Rodr{\'{i}}guez},\ and\ \citenamefont
  {Olaizola}}]{San-Blas_LIPSS_20}%
  \BibitemOpen
  \bibfield  {author} {\bibinfo {author} {\bibfnamefont {A.}~\bibnamefont
  {San-Blas}}, \bibinfo {author} {\bibfnamefont {M.}~\bibnamefont
  {Martinez-Calderon}}, \bibinfo {author} {\bibfnamefont {J.}~\bibnamefont
  {Buencuerpo}}, \bibinfo {author} {\bibfnamefont {L.~M.}\ \bibnamefont
  {Sanchez-Brea}}, \bibinfo {author} {\bibfnamefont {J.}~\bibnamefont {del
  Hoyo}}, \bibinfo {author} {\bibfnamefont {M.}~\bibnamefont
  {G{\'{o}}mez-Aranzadi}}, \bibinfo {author} {\bibfnamefont {A.}~\bibnamefont
  {Rodr{\'{i}}guez}}, \ and\ \bibinfo {author} {\bibfnamefont {S.~M.}\
  \bibnamefont {Olaizola}},\ }\bibfield  {title} {\enquote {\bibinfo {title}
  {{Femtosecond laser fabrication of LIPSS-based waveplates on metallic
  surfaces}},}\ }\href {\doibase 10.1016/j.apsusc.2020.146328} {\bibfield
  {journal} {\bibinfo  {journal} {Applied Surface Science}\ }\textbf {\bibinfo
  {volume} {520}},\ \bibinfo {pages} {146328} (\bibinfo {year}
  {2020})}\BibitemShut {NoStop}%
\bibitem [{\citenamefont {Li}\ \emph {et~al.}(2020)\citenamefont {Li},
  \citenamefont {Duan}, \citenamefont {Yi}, \citenamefont {Wang}, \citenamefont
  {Radjenovic}, \citenamefont {Tian},\ and\ \citenamefont {Li}}]{Li_LIPSS_20}%
  \BibitemOpen
  \bibfield  {author} {\bibinfo {author} {\bibfnamefont {Chao~Yu}\ \bibnamefont
  {Li}}, \bibinfo {author} {\bibfnamefont {Sai}\ \bibnamefont {Duan}}, \bibinfo
  {author} {\bibfnamefont {Jun}\ \bibnamefont {Yi}}, \bibinfo {author}
  {\bibfnamefont {Chen}\ \bibnamefont {Wang}}, \bibinfo {author} {\bibfnamefont
  {Petar~M.}\ \bibnamefont {Radjenovic}}, \bibinfo {author} {\bibfnamefont
  {Zhong~Qun}\ \bibnamefont {Tian}}, \ and\ \bibinfo {author} {\bibfnamefont
  {Jian~Feng}\ \bibnamefont {Li}},\ }\bibfield  {title} {\enquote {\bibinfo
  {title} {{Real-time detection of single-molecule reaction by plasmon-enhanced
  spectroscopy}},}\ }\href {\doibase 10.1126/sciadv.aba6012} {\bibfield
  {journal} {\bibinfo  {journal} {Science Advances}\ }\textbf {\bibinfo
  {volume} {6}},\ \bibinfo {pages} {1--9} (\bibinfo {year} {2020})}\BibitemShut
  {NoStop}%
\bibitem [{\citenamefont {Bonse}\ \emph {et~al.}(2018)\citenamefont {Bonse},
  \citenamefont {Kirner}, \citenamefont {Griepentrog}, \citenamefont
  {Spaltmann},\ and\ \citenamefont {Kr{\"{u}}ger}}]{Bonse_Tribology18}%
  \BibitemOpen
  \bibfield  {author} {\bibinfo {author} {\bibfnamefont {J{\"{o}}rn}\
  \bibnamefont {Bonse}}, \bibinfo {author} {\bibfnamefont {Sabrina~V.}\
  \bibnamefont {Kirner}}, \bibinfo {author} {\bibfnamefont {Michael}\
  \bibnamefont {Griepentrog}}, \bibinfo {author} {\bibfnamefont {Dirk}\
  \bibnamefont {Spaltmann}}, \ and\ \bibinfo {author} {\bibfnamefont
  {J{\"{o}}rg}\ \bibnamefont {Kr{\"{u}}ger}},\ }\bibfield  {title} {\enquote
  {\bibinfo {title} {{Femtosecond laser texturing of surfaces for tribological
  applications}},}\ }\href {\doibase 10.3390/ma11050801} {\bibfield  {journal}
  {\bibinfo  {journal} {Materials}\ }\textbf {\bibinfo {volume} {11}},\
  \bibinfo {pages} {1--19} (\bibinfo {year} {2018})}\BibitemShut {NoStop}%
\bibitem [{\citenamefont {Rudenko}\ \emph {et~al.}(2020)\citenamefont
  {Rudenko}, \citenamefont {Abou-Saleh}, \citenamefont {Pigeon}, \citenamefont
  {Mauclair}, \citenamefont {Garrelie}, \citenamefont {Stoian},\ and\
  \citenamefont {Colombier}}]{Rudenko20}%
  \BibitemOpen
  \bibfield  {author} {\bibinfo {author} {\bibfnamefont {A.}~\bibnamefont
  {Rudenko}}, \bibinfo {author} {\bibfnamefont {A.}~\bibnamefont {Abou-Saleh}},
  \bibinfo {author} {\bibfnamefont {F.}~\bibnamefont {Pigeon}}, \bibinfo
  {author} {\bibfnamefont {C.}~\bibnamefont {Mauclair}}, \bibinfo {author}
  {\bibfnamefont {F.}~\bibnamefont {Garrelie}}, \bibinfo {author}
  {\bibfnamefont {R.}~\bibnamefont {Stoian}}, \ and\ \bibinfo {author}
  {\bibfnamefont {J.P.}\ \bibnamefont {Colombier}},\ }\bibfield  {title}
  {\enquote {\bibinfo {title} {High-frequency periodic patterns driven by
  non-radiative fields coupled with marangoni convection instabilities on
  laser-excited metal surfaces},}\ }\href {\doibase
  https://doi.org/10.1016/j.actamat.2020.04.058} {\bibfield  {journal}
  {\bibinfo  {journal} {Acta Materialia}\ }\textbf {\bibinfo {volume} {194}},\
  \bibinfo {pages} {93--105} (\bibinfo {year} {2020})}\BibitemShut {NoStop}%
\bibitem [{\citenamefont {Roth}(2016)}]{Roth16}%
  \BibitemOpen
  \bibfield  {author} {\bibinfo {author} {\bibfnamefont {Stephan~V}\
  \bibnamefont {Roth}},\ }\bibfield  {title} {\enquote {\bibinfo {title} {A
  deep look into the spray coating process in real-time—the crucial role of
  x-rays},}\ }\href {\doibase 10.1088/0953-8984/28/40/403003} {\bibfield
  {journal} {\bibinfo  {journal} {Journal of Physics: Condensed Matter}\
  }\textbf {\bibinfo {volume} {28}},\ \bibinfo {pages} {403003} (\bibinfo
  {year} {2016})}\BibitemShut {NoStop}%
\bibitem [{\citenamefont {Su}\ \emph {et~al.}(2017)\citenamefont {Su},
  \citenamefont {Körstgens}, \citenamefont {Yao}, \citenamefont {Magerl},
  \citenamefont {Song}, \citenamefont {Metwalli}, \citenamefont {Bernstorff},\
  and\ \citenamefont {Müller-Buschbaum}}]{Su17}%
  \BibitemOpen
  \bibfield  {author} {\bibinfo {author} {\bibfnamefont {Bo}~\bibnamefont
  {Su}}, \bibinfo {author} {\bibfnamefont {Volker}\ \bibnamefont {Körstgens}},
  \bibinfo {author} {\bibfnamefont {Yuan}\ \bibnamefont {Yao}}, \bibinfo
  {author} {\bibfnamefont {David}\ \bibnamefont {Magerl}}, \bibinfo {author}
  {\bibfnamefont {Lin}\ \bibnamefont {Song}}, \bibinfo {author} {\bibfnamefont
  {Ezzeldin}\ \bibnamefont {Metwalli}}, \bibinfo {author} {\bibfnamefont
  {Sigrid}\ \bibnamefont {Bernstorff}}, \ and\ \bibinfo {author} {\bibfnamefont
  {Peter}\ \bibnamefont {Müller-Buschbaum}},\ }\bibfield  {title} {\enquote
  {\bibinfo {title} {Pore size control of block copolymer-templated
  sol–gel-synthesized titania films deposited via spray coating},}\ }\href
  {\doibase 10.1007/s10971-016-4134-9} {\bibfield  {journal} {\bibinfo
  {journal} {Journal of Sol-Gel Science and Technology}\ }\textbf {\bibinfo
  {volume} {81}},\ \bibinfo {pages} {346--354} (\bibinfo {year}
  {2017})}\BibitemShut {NoStop}%
\bibitem [{\citenamefont {Randolph}\ \emph {et~al.}(2022)\citenamefont
  {Randolph}, \citenamefont {Banjafar}, \citenamefont {Preston}, \citenamefont
  {Yabuuchi}, \citenamefont {Makita}, \citenamefont {Dover}, \citenamefont
  {R\"odel}, \citenamefont {G\"ode}, \citenamefont {Inubushi}, \citenamefont
  {Jakob}, \citenamefont {Kaa}, \citenamefont {Kon}, \citenamefont {Koga},
  \citenamefont {Ksenzov}, \citenamefont {Matsuoka}, \citenamefont {Nishiuchi},
  \citenamefont {Paulus}, \citenamefont {Schon}, \citenamefont {Sueda},
  \citenamefont {Sentoku}, \citenamefont {Togashi}, \citenamefont {Bussmann},
  \citenamefont {Cowan}, \citenamefont {Kl\"aui}, \citenamefont
  {Fortmann-Grote}, \citenamefont {Huang}, \citenamefont {Mancuso},
  \citenamefont {Kluge}, \citenamefont {Gutt},\ and\ \citenamefont
  {Nakatsutsumi}}]{Randolph22}%
  \BibitemOpen
  \bibfield  {author} {\bibinfo {author} {\bibfnamefont {L.}~\bibnamefont
  {Randolph}}, \bibinfo {author} {\bibfnamefont {M.}~\bibnamefont {Banjafar}},
  \bibinfo {author} {\bibfnamefont {T.~R.}\ \bibnamefont {Preston}}, \bibinfo
  {author} {\bibfnamefont {T.}~\bibnamefont {Yabuuchi}}, \bibinfo {author}
  {\bibfnamefont {M.}~\bibnamefont {Makita}}, \bibinfo {author} {\bibfnamefont
  {N.~P.}\ \bibnamefont {Dover}}, \bibinfo {author} {\bibfnamefont
  {C.}~\bibnamefont {R\"odel}}, \bibinfo {author} {\bibfnamefont
  {S.}~\bibnamefont {G\"ode}}, \bibinfo {author} {\bibfnamefont
  {Y.}~\bibnamefont {Inubushi}}, \bibinfo {author} {\bibfnamefont
  {G.}~\bibnamefont {Jakob}}, \bibinfo {author} {\bibfnamefont
  {J.}~\bibnamefont {Kaa}}, \bibinfo {author} {\bibfnamefont {A.}~\bibnamefont
  {Kon}}, \bibinfo {author} {\bibfnamefont {J.~K.}\ \bibnamefont {Koga}},
  \bibinfo {author} {\bibfnamefont {D.}~\bibnamefont {Ksenzov}}, \bibinfo
  {author} {\bibfnamefont {T.}~\bibnamefont {Matsuoka}}, \bibinfo {author}
  {\bibfnamefont {M.}~\bibnamefont {Nishiuchi}}, \bibinfo {author}
  {\bibfnamefont {M.}~\bibnamefont {Paulus}}, \bibinfo {author} {\bibfnamefont
  {F.}~\bibnamefont {Schon}}, \bibinfo {author} {\bibfnamefont
  {K.}~\bibnamefont {Sueda}}, \bibinfo {author} {\bibfnamefont
  {Y.}~\bibnamefont {Sentoku}}, \bibinfo {author} {\bibfnamefont
  {T.}~\bibnamefont {Togashi}}, \bibinfo {author} {\bibfnamefont
  {M.}~\bibnamefont {Bussmann}}, \bibinfo {author} {\bibfnamefont {T.~E.}\
  \bibnamefont {Cowan}}, \bibinfo {author} {\bibfnamefont {M.}~\bibnamefont
  {Kl\"aui}}, \bibinfo {author} {\bibfnamefont {C.}~\bibnamefont
  {Fortmann-Grote}}, \bibinfo {author} {\bibfnamefont {L.}~\bibnamefont
  {Huang}}, \bibinfo {author} {\bibfnamefont {A.~P.}\ \bibnamefont {Mancuso}},
  \bibinfo {author} {\bibfnamefont {T.}~\bibnamefont {Kluge}}, \bibinfo
  {author} {\bibfnamefont {C.}~\bibnamefont {Gutt}}, \ and\ \bibinfo {author}
  {\bibfnamefont {M.}~\bibnamefont {Nakatsutsumi}},\ }\bibfield  {title}
  {\enquote {\bibinfo {title} {Nanoscale subsurface dynamics of solids upon
  high-intensity femtosecond laser irradiation observed by grazing-incidence
  x-ray scattering},}\ }\href {\doibase 10.1103/PhysRevResearch.4.033038}
  {\bibfield  {journal} {\bibinfo  {journal} {Phys. Rev. Research}\ }\textbf
  {\bibinfo {volume} {4}},\ \bibinfo {pages} {033038} (\bibinfo {year}
  {2022})}\BibitemShut {NoStop}%
\bibitem [{\citenamefont {Hol\'{y}}\ \emph {et~al.}(1993)\citenamefont
  {Hol\'{y}}, \citenamefont {Kub\v{e}na}, \citenamefont {Ohl\'{i}dal},
  \citenamefont {Lischka},\ and\ \citenamefont {Plotz}}]{Holy93}%
  \BibitemOpen
  \bibfield  {author} {\bibinfo {author} {\bibfnamefont {V.}~\bibnamefont
  {Hol\'{y}}}, \bibinfo {author} {\bibfnamefont {J.}~\bibnamefont
  {Kub\v{e}na}}, \bibinfo {author} {\bibfnamefont {I.}~\bibnamefont
  {Ohl\'{i}dal}}, \bibinfo {author} {\bibfnamefont {K.}~\bibnamefont
  {Lischka}}, \ and\ \bibinfo {author} {\bibfnamefont {W.}~\bibnamefont
  {Plotz}},\ }\bibfield  {title} {\enquote {\bibinfo {title} {{X-ray reflection
  from rough layered systems}},}\ }\href {\doibase 10.1103/PhysRevB.47.15896}
  {\bibfield  {journal} {\bibinfo  {journal} {Phys. Rev. B}\ }\textbf {\bibinfo
  {volume} {47}},\ \bibinfo {pages} {15896--15903} (\bibinfo {year}
  {1993})}\BibitemShut {NoStop}%
\bibitem [{\citenamefont {Sinha}\ \emph {et~al.}(1988)\citenamefont {Sinha},
  \citenamefont {Sirota}, \citenamefont {Garoff},\ and\ \citenamefont
  {Stanley}}]{PhysRevB.38.2297}%
  \BibitemOpen
  \bibfield  {author} {\bibinfo {author} {\bibfnamefont {S.~K.}\ \bibnamefont
  {Sinha}}, \bibinfo {author} {\bibfnamefont {E.~B.}\ \bibnamefont {Sirota}},
  \bibinfo {author} {\bibfnamefont {S.}~\bibnamefont {Garoff}}, \ and\ \bibinfo
  {author} {\bibfnamefont {H.~B.}\ \bibnamefont {Stanley}},\ }\bibfield
  {title} {\enquote {\bibinfo {title} {X-ray and neutron scattering from rough
  surfaces},}\ }\href {\doibase 10.1103/PhysRevB.38.2297} {\bibfield  {journal}
  {\bibinfo  {journal} {Phys. Rev. B}\ }\textbf {\bibinfo {volume} {38}},\
  \bibinfo {pages} {2297--2311} (\bibinfo {year} {1988})}\BibitemShut {NoStop}%
\bibitem [{\citenamefont {Schlomka}\ \emph {et~al.}(1995)\citenamefont
  {Schlomka}, \citenamefont {Tolan}, \citenamefont {Schwalowsky}, \citenamefont
  {Seeck}, \citenamefont {Stettner},\ and\ \citenamefont
  {Press}}]{PhysRevB.51.2311}%
  \BibitemOpen
  \bibfield  {author} {\bibinfo {author} {\bibfnamefont {J.-P.}\ \bibnamefont
  {Schlomka}}, \bibinfo {author} {\bibfnamefont {M.}~\bibnamefont {Tolan}},
  \bibinfo {author} {\bibfnamefont {L.}~\bibnamefont {Schwalowsky}}, \bibinfo
  {author} {\bibfnamefont {O.~H.}\ \bibnamefont {Seeck}}, \bibinfo {author}
  {\bibfnamefont {J.}~\bibnamefont {Stettner}}, \ and\ \bibinfo {author}
  {\bibfnamefont {W.}~\bibnamefont {Press}},\ }\bibfield  {title} {\enquote
  {\bibinfo {title} {X-ray diffraction from si/ge layers: Diffuse scattering in
  the region of total external reflection},}\ }\href {\doibase
  10.1103/PhysRevB.51.2311} {\bibfield  {journal} {\bibinfo  {journal} {Phys.
  Rev. B}\ }\textbf {\bibinfo {volume} {51}},\ \bibinfo {pages} {2311--2321}
  (\bibinfo {year} {1995})}\BibitemShut {NoStop}%
\bibitem [{\citenamefont {Yabuuchi}\ \emph {et~al.}(2019)\citenamefont
  {Yabuuchi}, \citenamefont {Kon}, \citenamefont {Inubushi}, \citenamefont
  {Togahi}, \citenamefont {Sueda}, \citenamefont {Itoga}, \citenamefont
  {Nakajima}, \citenamefont {Habara}, \citenamefont {Kodama}, \citenamefont
  {Tomizawa},\ and\ \citenamefont {Yabashi}}]{Yabuuchi19}%
  \BibitemOpen
  \bibfield  {author} {\bibinfo {author} {\bibfnamefont {T.}~\bibnamefont
  {Yabuuchi}}, \bibinfo {author} {\bibfnamefont {A.}~\bibnamefont {Kon}},
  \bibinfo {author} {\bibfnamefont {Y.}~\bibnamefont {Inubushi}}, \bibinfo
  {author} {\bibfnamefont {T.}~\bibnamefont {Togahi}}, \bibinfo {author}
  {\bibfnamefont {K.}~\bibnamefont {Sueda}}, \bibinfo {author} {\bibfnamefont
  {T.}~\bibnamefont {Itoga}}, \bibinfo {author} {\bibfnamefont
  {K.}~\bibnamefont {Nakajima}}, \bibinfo {author} {\bibfnamefont
  {H.}~\bibnamefont {Habara}}, \bibinfo {author} {\bibfnamefont
  {R.}~\bibnamefont {Kodama}}, \bibinfo {author} {\bibfnamefont
  {H.}~\bibnamefont {Tomizawa}}, \ and\ \bibinfo {author} {\bibfnamefont
  {M.}~\bibnamefont {Yabashi}},\ }\bibfield  {title} {\enquote {\bibinfo
  {title} {{An experimental platform using high-power, high-intensity optical
  lasers with the hard X-ray free-electron laser at SACLA}},}\ }\href {\doibase
  10.1107/S1600577519000882} {\bibfield  {journal} {\bibinfo  {journal} {J.
  Sync. Rad.}\ }\textbf {\bibinfo {volume} {26}},\ \bibinfo {pages} {585--594}
  (\bibinfo {year} {2019})}\BibitemShut {NoStop}%
\bibitem [{\citenamefont {Kameshima}\ \emph {et~al.}(2014)\citenamefont
  {Kameshima}, \citenamefont {Ono}, \citenamefont {Kudo}, \citenamefont
  {Ozaki}, \citenamefont {Kirihara}, \citenamefont {Kobayashi}, \citenamefont
  {Inubushi}, \citenamefont {Yabashi}, \citenamefont {Horigome}, \citenamefont
  {Holland}, \citenamefont {Holland}, \citenamefont {Burt}, \citenamefont
  {Murao},\ and\ \citenamefont {Hatsui}}]{Kameshima14}%
  \BibitemOpen
  \bibfield  {author} {\bibinfo {author} {\bibfnamefont {T.}~\bibnamefont
  {Kameshima}}, \bibinfo {author} {\bibfnamefont {S.}~\bibnamefont {Ono}},
  \bibinfo {author} {\bibfnamefont {T.}~\bibnamefont {Kudo}}, \bibinfo {author}
  {\bibfnamefont {K.}~\bibnamefont {Ozaki}}, \bibinfo {author} {\bibfnamefont
  {Y.}~\bibnamefont {Kirihara}}, \bibinfo {author} {\bibfnamefont
  {K.}~\bibnamefont {Kobayashi}}, \bibinfo {author} {\bibfnamefont
  {Y.}~\bibnamefont {Inubushi}}, \bibinfo {author} {\bibfnamefont
  {M.}~\bibnamefont {Yabashi}}, \bibinfo {author} {\bibfnamefont
  {T.}~\bibnamefont {Horigome}}, \bibinfo {author} {\bibfnamefont
  {A.}~\bibnamefont {Holland}}, \bibinfo {author} {\bibfnamefont
  {K.}~\bibnamefont {Holland}}, \bibinfo {author} {\bibfnamefont
  {D.}~\bibnamefont {Burt}}, \bibinfo {author} {\bibfnamefont {H.}~\bibnamefont
  {Murao}}, \ and\ \bibinfo {author} {\bibfnamefont {T.}~\bibnamefont
  {Hatsui}},\ }\bibfield  {title} {\enquote {\bibinfo {title} {{Development of
  an X-ray pixel detector with multi-port charge-coupled device for X-ray
  free-electron laser experiments}},}\ }\href {\doibase 10.1063/1.4867668}
  {\bibfield  {journal} {\bibinfo  {journal} {Review of Scientific
  Instruments}\ }\textbf {\bibinfo {volume} {85}},\ \bibinfo {pages} {033110}
  (\bibinfo {year} {2014})}\BibitemShut {NoStop}%
\bibitem [{\citenamefont {Kiessig}(1931)}]{Kiessig31}%
  \BibitemOpen
  \bibfield  {author} {\bibinfo {author} {\bibfnamefont {H.}~\bibnamefont
  {Kiessig}},\ }\bibfield  {title} {\enquote {\bibinfo {title} {{Untersuchungen
  zur Totalreflexion von R\"{o}ntgenstrahlen}},}\ }\href@noop {} {\bibfield
  {journal} {\bibinfo  {journal} {Annalen der Physik}\ }\textbf {\bibinfo
  {volume} {10}},\ \bibinfo {pages} {769} (\bibinfo {year} {1931})}\BibitemShut
  {NoStop}%
\bibitem [{\citenamefont {Yoneda}(1963)}]{Yoneda63}%
  \BibitemOpen
  \bibfield  {author} {\bibinfo {author} {\bibfnamefont {Y.}~\bibnamefont
  {Yoneda}},\ }\bibfield  {title} {\enquote {\bibinfo {title} {{Anomalous
  Surface Reflection of X Rays}},}\ }\href {\doibase 10.1103/PhysRev.131.2010}
  {\bibfield  {journal} {\bibinfo  {journal} {Phys. Rev.}\ }\textbf {\bibinfo
  {volume} {131}},\ \bibinfo {pages} {2010--2013} (\bibinfo {year}
  {1963})}\BibitemShut {NoStop}%
\bibitem [{\citenamefont {Pospelov}\ \emph {et~al.}(2020)\citenamefont
  {Pospelov}, \citenamefont {Van~Herck}, \citenamefont {Burle}, \citenamefont
  {Carmona~Loaiza}, \citenamefont {Durniak}, \citenamefont {Fisher},
  \citenamefont {Ganeva}, \citenamefont {Yurov},\ and\ \citenamefont
  {Wuttke}}]{Pospelov:ge5067}%
  \BibitemOpen
  \bibfield  {author} {\bibinfo {author} {\bibfnamefont {Gennady}\ \bibnamefont
  {Pospelov}}, \bibinfo {author} {\bibfnamefont {Walter}\ \bibnamefont
  {Van~Herck}}, \bibinfo {author} {\bibfnamefont {Jan}\ \bibnamefont {Burle}},
  \bibinfo {author} {\bibfnamefont {Juan~M.}\ \bibnamefont {Carmona~Loaiza}},
  \bibinfo {author} {\bibfnamefont {C{\'{e}}line}\ \bibnamefont {Durniak}},
  \bibinfo {author} {\bibfnamefont {Jonathan~M.}\ \bibnamefont {Fisher}},
  \bibinfo {author} {\bibfnamefont {Marina}\ \bibnamefont {Ganeva}}, \bibinfo
  {author} {\bibfnamefont {Dmitry}\ \bibnamefont {Yurov}}, \ and\ \bibinfo
  {author} {\bibfnamefont {Joachim}\ \bibnamefont {Wuttke}},\ }\bibfield
  {title} {\enquote {\bibinfo {title} {{{\it BornAgain}: software for
  simulating and fitting grazing-incidence small-angle scattering}},}\ }\href
  {\doibase 10.1107/S1600576719016789} {\bibfield  {journal} {\bibinfo
  {journal} {Journal of Applied Crystallography}\ }\textbf {\bibinfo {volume}
  {53}},\ \bibinfo {pages} {262--276} (\bibinfo {year} {2020})}\BibitemShut
  {NoStop}%
\bibitem [{\citenamefont {Salditt}\ \emph {et~al.}(1994)\citenamefont
  {Salditt}, \citenamefont {Metzger},\ and\ \citenamefont
  {Peisl}}]{PhysRevLett.73.2228}%
  \BibitemOpen
  \bibfield  {author} {\bibinfo {author} {\bibfnamefont {T.}~\bibnamefont
  {Salditt}}, \bibinfo {author} {\bibfnamefont {T.~H.}\ \bibnamefont
  {Metzger}}, \ and\ \bibinfo {author} {\bibfnamefont {J.}~\bibnamefont
  {Peisl}},\ }\bibfield  {title} {\enquote {\bibinfo {title} {Kinetic roughness
  of amorphous multilayers studied by diffuse x-ray scattering},}\ }\href
  {\doibase 10.1103/PhysRevLett.73.2228} {\bibfield  {journal} {\bibinfo
  {journal} {Phys. Rev. Lett.}\ }\textbf {\bibinfo {volume} {73}},\ \bibinfo
  {pages} {2228--2231} (\bibinfo {year} {1994})}\BibitemShut {NoStop}%
\bibitem [{\citenamefont {Nguyen}\ \emph {et~al.}(2018)\citenamefont {Nguyen},
  \citenamefont {Hashimoto}, \citenamefont {Zakharov}, \citenamefont {Stach},
  \citenamefont {Rooney}, \citenamefont {Berkels}, \citenamefont {Thompson},
  \citenamefont {Haigh},\ and\ \citenamefont
  {Burnett}}]{doi:10.1021/acsami.7b17224}%
  \BibitemOpen
  \bibfield  {author} {\bibinfo {author} {\bibfnamefont {Lan}\ \bibnamefont
  {Nguyen}}, \bibinfo {author} {\bibfnamefont {Teruo}\ \bibnamefont
  {Hashimoto}}, \bibinfo {author} {\bibfnamefont {Dmitri~N.}\ \bibnamefont
  {Zakharov}}, \bibinfo {author} {\bibfnamefont {Eric~A.}\ \bibnamefont
  {Stach}}, \bibinfo {author} {\bibfnamefont {Aidan~P.}\ \bibnamefont
  {Rooney}}, \bibinfo {author} {\bibfnamefont {Benjamin}\ \bibnamefont
  {Berkels}}, \bibinfo {author} {\bibfnamefont {George~E.}\ \bibnamefont
  {Thompson}}, \bibinfo {author} {\bibfnamefont {Sarah~J.}\ \bibnamefont
  {Haigh}}, \ and\ \bibinfo {author} {\bibfnamefont {Tim~L.}\ \bibnamefont
  {Burnett}},\ }\bibfield  {title} {\enquote {\bibinfo {title} {Atomic-scale
  insights into the oxidation of aluminum},}\ }\href {\doibase
  10.1021/acsami.7b17224} {\bibfield  {journal} {\bibinfo  {journal} {ACS
  Applied Materials \& Interfaces}\ }\textbf {\bibinfo {volume} {10}},\
  \bibinfo {pages} {2230--2235} (\bibinfo {year} {2018})},\ \bibinfo {note}
  {pMID: 29319290},\ \Eprint
  {http://arxiv.org/abs/https://doi.org/10.1021/acsami.7b17224}
  {https://doi.org/10.1021/acsami.7b17224} \BibitemShut {NoStop}%
\bibitem [{\citenamefont {Sentoku}\ and\ \citenamefont
  {Kemp}(2008)}]{SENTOKU2008}%
  \BibitemOpen
  \bibfield  {author} {\bibinfo {author} {\bibfnamefont {Y.}~\bibnamefont
  {Sentoku}}\ and\ \bibinfo {author} {\bibfnamefont {A.J.}\ \bibnamefont
  {Kemp}},\ }\bibfield  {title} {\enquote {\bibinfo {title} {Numerical methods
  for particle simulations at extreme densities and temperatures: Weighted
  particles, relativistic collisions and reduced currents},}\ }\href {\doibase
  https://doi.org/10.1016/j.jcp.2008.03.043} {\bibfield  {journal} {\bibinfo
  {journal} {Journal of Computational Physics}\ }\textbf {\bibinfo {volume}
  {227}},\ \bibinfo {pages} {6846 -- 6861} (\bibinfo {year}
  {2008})}\BibitemShut {NoStop}%
\bibitem [{\citenamefont {Paschke-Bruehl}\ \emph {et~al.}(2023)\citenamefont
  {Paschke-Bruehl}, \citenamefont {Banjafar}, \citenamefont {Garten},
  \citenamefont {Huang}, \citenamefont {Marré}, \citenamefont {Nakatsutsumi},
  \citenamefont {Randolph}, \citenamefont {Cowan}, \citenamefont {Schramm},\
  and\ \citenamefont {Kluge}}]{Paschke2022}%
  \BibitemOpen
  \bibfield  {author} {\bibinfo {author} {\bibfnamefont {Franziska-Luise}\
  \bibnamefont {Paschke-Bruehl}}, \bibinfo {author} {\bibfnamefont
  {Mohammadreza}\ \bibnamefont {Banjafar}}, \bibinfo {author} {\bibfnamefont
  {Marco}\ \bibnamefont {Garten}}, \bibinfo {author} {\bibfnamefont {Lingen}\
  \bibnamefont {Huang}}, \bibinfo {author} {\bibfnamefont {Brian~Edward}\
  \bibnamefont {Marré}}, \bibinfo {author} {\bibfnamefont {Motoaki}\
  \bibnamefont {Nakatsutsumi}}, \bibinfo {author} {\bibfnamefont {Lisa}\
  \bibnamefont {Randolph}}, \bibinfo {author} {\bibfnamefont {Thomas~E.}\
  \bibnamefont {Cowan}}, \bibinfo {author} {\bibfnamefont {Ulrich}\
  \bibnamefont {Schramm}}, \ and\ \bibinfo {author} {\bibfnamefont {Thomas}\
  \bibnamefont {Kluge}},\ }\bibfield  {title} {\enquote {\bibinfo {title}
  {Heating in multi-layer targets at ultra-high intensity laser irradiation and
  the impact of density oscillation},}\ }\href@noop {} {\bibfield  {journal}
  {\bibinfo  {journal} {New J. Phys.}\ }\textbf {\bibinfo {volume} {25}},\
  \bibinfo {pages} {043024} (\bibinfo {year} {2023})}\BibitemShut {NoStop}%
\bibitem [{\citenamefont {Eidmann}\ \emph {et~al.}(2000)\citenamefont
  {Eidmann}, \citenamefont {Meyer-ter Vehn}, \citenamefont {Schlegel},\ and\
  \citenamefont {H\"uller}}]{Eidmann00}%
  \BibitemOpen
  \bibfield  {author} {\bibinfo {author} {\bibfnamefont {K.}~\bibnamefont
  {Eidmann}}, \bibinfo {author} {\bibfnamefont {J.}~\bibnamefont {Meyer-ter
  Vehn}}, \bibinfo {author} {\bibfnamefont {T.}~\bibnamefont {Schlegel}}, \
  and\ \bibinfo {author} {\bibfnamefont {S.}~\bibnamefont {H\"uller}},\
  }\bibfield  {title} {\enquote {\bibinfo {title} {Hydrodynamic simulation of
  subpicosecond laser interaction with solid-density matter},}\ }\href
  {\doibase 10.1103/PhysRevE.62.1202} {\bibfield  {journal} {\bibinfo
  {journal} {Phys. Rev. E}\ }\textbf {\bibinfo {volume} {62}},\ \bibinfo
  {pages} {1202--1214} (\bibinfo {year} {2000})}\BibitemShut {NoStop}%
\bibitem [{\citenamefont {Ramis}\ \emph {et~al.}(2012)\citenamefont {Ramis},
  \citenamefont {Eidmann}, \citenamefont {Meyer-ter Vehn},\ and\ \citenamefont
  {H{\"u}ller}}]{Ramis2012}%
  \BibitemOpen
  \bibfield  {author} {\bibinfo {author} {\bibfnamefont {R}~\bibnamefont
  {Ramis}}, \bibinfo {author} {\bibfnamefont {K}~\bibnamefont {Eidmann}},
  \bibinfo {author} {\bibfnamefont {J}~\bibnamefont {Meyer-ter Vehn}}, \ and\
  \bibinfo {author} {\bibfnamefont {S}~\bibnamefont {H{\"u}ller}},\ }\bibfield
  {title} {\enquote {\bibinfo {title} {Multi-fs--a computer code for
  laser--plasma interaction in the femtosecond regime},}\ }\href@noop {}
  {\bibfield  {journal} {\bibinfo  {journal} {Computer Physics Communications}\
  }\textbf {\bibinfo {volume} {183}},\ \bibinfo {pages} {637--655} (\bibinfo
  {year} {2012})}\BibitemShut {NoStop}%
\bibitem [{\citenamefont {Faik}\ \emph {et~al.}(2018)\citenamefont {Faik},
  \citenamefont {Tauschwitz},\ and\ \citenamefont {Iosilevskiy}}]{Faik18}%
  \BibitemOpen
  \bibfield  {author} {\bibinfo {author} {\bibfnamefont {S.}~\bibnamefont
  {Faik}}, \bibinfo {author} {\bibfnamefont {A.}~\bibnamefont {Tauschwitz}}, \
  and\ \bibinfo {author} {\bibfnamefont {I.}~\bibnamefont {Iosilevskiy}},\
  }\bibfield  {title} {\enquote {\bibinfo {title} {{The equation of state
  package FEOS for high energy density matter}},}\ }\href {\doibase
  https://doi.org/10.1016/j.cpc.2018.01.008} {\bibfield  {journal} {\bibinfo
  {journal} {Comput. Phys. Comm.}\ }\textbf {\bibinfo {volume} {227}},\
  \bibinfo {pages} {117 -- 125} (\bibinfo {year} {2018})}\BibitemShut {NoStop}%
\end{thebibliography}%

\end{document}